\documentclass[aps,reprint]{revtex4-1}

\usepackage{graphics}
\usepackage[pdftex]{graphicx}

\begin{document}

\title{Ab initio simulations of hydrogen-bonded ferroelectrics: collective tunneling and the origin of geometrical isotope effects}

\author{K. T. Wikfeldt}
\email[]{wikfeldt@hi.is}
\affiliation{Science Institute, University of Iceland, Nordita, Stockholm, Sweden and University College London, London WC1E 6BT, United Kingdom}
\author{A. Michaelides}
\affiliation{Thomas Young Centre, London Centre for Nanotechnology and Department of Chemistry, University College London, London WC1E 6BT, United Kingdom}

\date{\today}

\begin{abstract}
Ab initio simulations that account for nuclear quantum effects have been 
used to examine the order-disorder transition in squaric acid, a 
prototypical H-bonded antiferroelectric crystal. Our simulations reproduce 
the $>$100 K difference in transition temperature observed upon deuteration as 
well as the strong geometrical isotope effect observed on intermolecular 
separations within the crystal. We find that collective transfer of protons 
along the H-bonding chains -- facilitated by quantum mechanical tunneling 
-- is critical to the order-disorder transition and the geometrical isotope 
effect. This sheds light on the origin of isotope effects and the importance 
of tunneling in squaric acid which likely extends to other H-bonded 
ferroelectrics.
\end{abstract}

\maketitle

Ferroelectric materials have been extensively examined because of their many 
diverse applications in, \textit{e.g.}, electro-optic, piezoelectric and 
random access memory devices~\cite{blinc2002ferroelectrics}. Recently 
interest has intensified in hydrogen (H-) bonded ferroelectrics because 
of the discovery of above room-temperature ferroelectricity in an organic 
crystal and the realization that they could potentially be used as 
cheaper and more environmentally friendly organic 
electronics~\cite{horiuchi2005ferro,horiuchi2008organic,horiuchi2010above}.  
H-bonded 
ferroelectrics are also a valuable class of materials through which we can 
gain deeper understanding of the fundamental nature of H-bonding. Primarily 
this is because they are well-characterized crystalline materials with a 
range of H-bonding configurations and in most cases have been synthesized in 
both their standard and deuterated forms. 

Many H-bonded ferroelectrics exhibit phase transitions to paraelectric phases 
that lack long-range ordering of protons in the H-bonds. The Curie temperature 
(T$_c$) of these transitions can dramatically increase by $\sim$100 K upon 
deuteration but the physical origin of this effect is still not fully understood. 
An early model~\cite{blinc1960isotopic,blinc1966cluster1} explained this giant isotope 
effect on the basis of tunneling of protons 
in double well potentials. However, this model fails to account for experimentally 
observed geometrical isotope effects in H-bonding geometry between the protonated and 
deuterated crystals, where H-bonds have been observed to elongate upon 
deuteration~\cite{mcmahon1990geometric,mcmahon1991effect} -- a so called Ubbelohde 
effect~\cite{ubbelohde1955acid,li2011quantum}.
Models involving a coupling between lattice modes and proton dynamics were therefore 
suggested~\cite{bussmann1998bond,dalal1998coexistence}.
It was argued on these grounds
that tunneling is unnecessary to explain the large increase of $T_c$  
upon deuteration~\cite{mcmahon1990geometric}. On the other hand, neutron Compton scattering 
experiments on the commonly studied KH$_2$PO$_4$ (KDP) 
system ~\cite{reiter2002direct} 
showed that protons 
occupy both sites along the H-bonds on a short time-scale above its phase transition at 124~K, 
indicating coherent quantum tunneling, while no such coherence was found in 
the deuterated crystal~\cite{reiter2008deuteron}. 
Theoretical work~\cite{koval2002ferroelectricity,koval2005first} 
attempted to reconcile these differing interpretations by suggesting that a mechanism 
behind the Ubbelohde effect may itself be collective tunneling in clusters of atoms 
in the crystal.
However, direct \textit{ab initio} simulations aimed at elucidating the mechanisms behind 
the isotope shift of $T_c$ and Ubbelohde effects are challenging due to the complex coupled dynamics 
involving multiple H/D atoms, and the role of tunneling has thus remained unclear.
In this context, squaric acid (C$_4$H$_2$O$_4$ or H$_2$SQ) provides 
an ideal simple model system which has been explored extensively 
by both experiments~\cite{petersson1981phase,katrusiak1986pressure,mcmahon1991effect,semmingsen1995ZfK,dalal1998coexistence} and theory~\cite{dolin2011quantum,rovira2001hydrogen,palomar2002GIAO,bussmann2007order,ishizuka2011quantum}. 

Here we report results from large-scale \textit{ab initio} path-integral molecular dynamics (PIMD) simulations 
of H$_2$SQ and its deuterated analogue D$_2$SQ that have allowed us to 
extract detailed information on quantum effects and collective proton behavior. 
\textit{Ab initio} PIMD was recently applied to 
KDP~\cite{srinivasan2011isotope}, but due to the limited system size 
where collective effects were neglected only the disordered paraelectric 
phase could be studied. We find that collective effects are vital for 
capturing antiferroelectric order in the simulations. Further, the Ubbelohde effect is 
well reproduced by PIMD and the transition to paraelectric ordering occurs at higher temperature 
for D$_2$SQ compared to H$_2$SQ, in agreement with the experimental shift from $T_c$(H$_2$SQ)=373~K
to $T_c$(D$_2$SQ)=520~K. 
Our simulations also reveal that tunneling is an important mechanism behind both the Ubbelohde effect and 
the collective proton jumps in the paraelectric phase.

\begin{figure}[h]
\centering
  \includegraphics[width=0.4\textwidth]{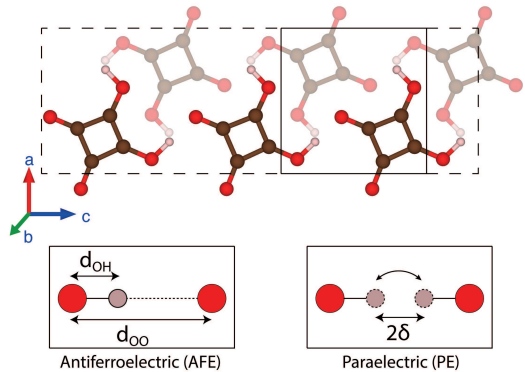}
  \caption{Crystal structure of squaric acid. The solid and dashed squares in the upper figure denote 
the primitive unit cell and the 1$\times$1$\times$3 supercell used for PIMD 
simulations, respectively. 
The lower panel illustrates the $d_{\mathrm{OO}}$ and $\delta$ structural parameters 
and the difference between AFE and PE ordering.}
  \label{fig:vis}
\end{figure}

Squaric acid is composed of planar C$_4$H$_2$O$_4$ molecules 
bound by strong H-bonds in 2D sheets with 
two slightly inequivalent H-bonding chains running along the $a$- and $c$-axes (see 
Fig.~\ref{fig:vis}). The sheets in turn are stacked along the $b$-axis and weakly 
bound by dispersion forces. In the low-temperature antiferroelectric (AFE) 
phase each layer is polarized due to long-range ordering of H atoms 
into equivalent sites in the H-bonds, with opposite polarization in 
neighboring planes, while in 
the high-temperature paraelectric (PE) phase the long-range ordering is 
lost due to proton disorder. 

Realistic simulations of H-bonded ferroelectrics are 
challenging for several reasons. First, an accurate theoretical approach is 
needed that describes the H-bonds between the molecules, the proton transfer 
barriers and, in many cases, weak van der Waals bonding between molecules
or layers of the material. 
Second, to capture the collective nature of proton ordering 
in squaric acid it is essential to use extended simulation cells with several 
molecular units along the H-bonding directions. Third, thermal and nuclear quantum 
effects such as tunneling and zero point motion must be taken into 
account given the finite temperature phase transition 
and the quantum mechanical nature of the proton. In tackling this system 
we considered all of these issues in detail. A more complete description of 
the theoretical approach employed and tests performed to establish its accuracy 
is given in the supplementary information~\cite{supmat}. 
In brief, the essential features of the simulations performed are that we 
used density functional theory (DFT) with projector-augmented wave (PAW) 
potentials~\cite{kresse1999ultrasoft} as implemented in the 
VASP code~\cite{kresse1996vasp1,kresse1996vasp2}. Nuclear quantum effects 
were accounted for with PIMD 
(see, \textit{e.g.}, refs~\cite{marx1994ab,tuckerman1997quantum} 
as implemented in VASP by Alf\`{e} and Gillan~\cite{alfe2010ab}). 32 path 
integral beads were used except for H$_2$SQ at 100-200 K where 64 beads were 
required and for D$_2$SQ at 300-500 K where 16 beads gave sufficiently converged 
quantum effects~\cite{supmat}. 
To capture collective effects we used 1$\times$1$\times$3 supercells 
(Fig.~\ref{fig:vis}) with a H-bonding 
chain of 3 molecular units along the $c$-axis and a total of 6 molecules 
in the cell. This setup also allowed us to investigate the role of 
collective effects by comparing the proton ordering along the $a$- and 
$c$-axes, which have been shown to be very weakly
coupled~\cite{ehrhardt1984one,palomar2002quantum}. 
All simulations were performed in the NVT 
ensemble with experimental lattice constants~\cite{supmat}. 
The vdW-DF2~\cite{lee2010vdwdf2,klimes2011van} 
exchange-correlation (xc) functional was selected after an extensive series 
of tests showed that of the xc functionals considered it gave good agreement 
with experiment in terms of O-O intermolecular distances in PIMD simulations 
and produced a proton 
transfer barrier in closest agreement with results from explicitly correlated 
calculations. 
Specifically, while the commonly used Perdew-Burke-Ernzerhof (PBE) functional 
gives a proton transfer barrier of 18~meV, the vdW-DF2 gives 87~meV which is 
considerably closer to values from M{\o}ller-Plesset second order perturbation theory (MP2, 110~meV) and random phase approximation (RPA, 148~meV) 
calculations~\cite{supmat}.

\begin{figure}[h]
\centering
  \includegraphics[width=0.45\textwidth]{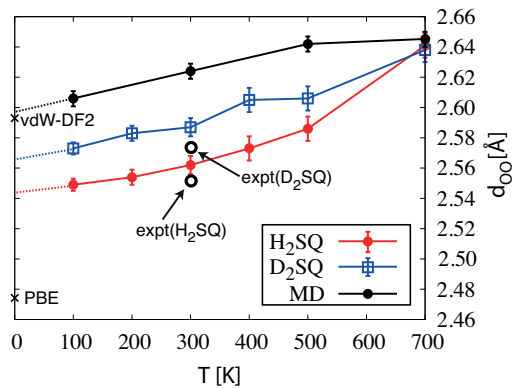}
  \caption{Average oxygen-oxygen distances $d_{\mathrm{OO}}$ as function of temperature for MD and PIMD simulations using vdW-DF2. $d_{\mathrm{OO}}$ values for optimized geometries using PBE and vdW-DF2 are shown by crosses at T=0. 
Experimental values for H$_2$SQ from ref.~\cite{semmingsen1977jcp} and D$_2$SQ 
from ref.~\cite{mcmahon1991effect} at 300~K are marked by empty circles.}
  \label{fig:func}
\end{figure}

We begin by discussing oxygen-oxygen distances, $d_{\mathrm{OO}}$, and their temperature and isotope dependence. 
Optimized $d_{\mathrm{OO}}$ values 
are shown by crosses in Fig.~\ref{fig:func} where it can be seen that 
vdW-DF2 predicts a value higher than experiment. 
Moreover, MD simulations give average $d_{\mathrm{OO}}$ values 
that increase with temperature, 
extrapolating down to the optimized value at 0~K and severely overestimating 
$d_{\mathrm{OO}}$ for H$_2$SQ by 0.07~\AA\ at 300~K compared to the 
experimental value.
This thermal exansion is however completely counteracted by quantum effects in 
PIMD, where at 300~K the combined thermal and quantum effects reduce 
$d_{\mathrm{OO}}$ by 0.06~\AA\ for H$_2$SQ and 0.04~\AA\ for D$_2$SQ compared 
to the optimized value, resulting in agreement 
within 0.01~\AA\ with the corresponding experimental 
values~\cite{semmingsen1977jcp,mcmahon1991effect}.
The large discrepancy between optimized and PIMD values for $d_{\mathrm{OO}}$ 
shows that geometry optimizations can be in serious error in systems with strong 
H-bonds and that quantum effects must be accounted for.
Note that the optimized PBE value severely underestimates the experimental 
$d_{\mathrm{OO}}$, an error that is further exasperated by the inclusion of 
quantum effects, highlighting the failure of the common PBE approximation 
to describe squaric acid. 
The observed Ubbelohde effect in PIMD, \textit{i.e.}, 
the difference in $d_{\mathrm{OO}}$ between 
H$_2$SQ and D$_2$SQ of around 0.02~\AA\ as seen in Fig.~\ref{fig:func}, 
is in good agreement with experiments. Starting 
at around 500~K this isotope effect decreases and only around 700~K (above the decomposition temperature 
of H$_2$SQ~\cite{lee2012polymorphic}) do the simulations predict the same 
$d_{\mathrm{OO}}$. MD also gives the same $d_{\mathrm{OO}}$ as the PIMD simulations at 700~K, 
consistent with vanishing influence of quantum effects at high temperatures. 

The $d_{\mathrm{OO}}$ values from MD and PIMD shown in Fig.~\ref{fig:func} are calculated along the $c$-axis 
where the coupling between three molecules is explicitly included. Interestingly, $d_{\mathrm{OO}}$ 
values along the $a$-axis where this coupling is neglected are shorter by around 0.02~\AA\ for both 
H$_2$SQ and D$_2$SQ. This shortening is connected to more frequent 
jumps of H and D atoms along the $a$-axis compared to the $c$-axis. Collective effects thus 
enhance the localization of H and D atoms resulting in longer H-bonds.

\begin{figure}[h]
\centering
  \includegraphics[width=0.5\textwidth]{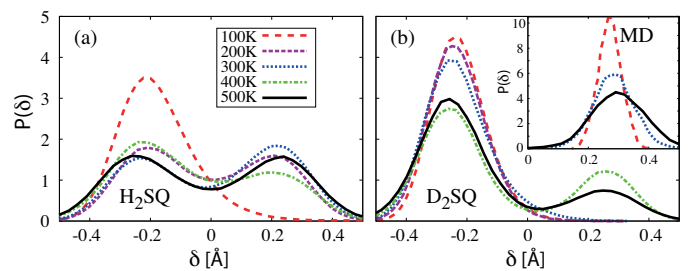}
  \caption{Probability distributions of the $\delta$ parameter for (a) H$_2$SQ and (b) D$_2$SQ from PIMD, and (inset in (b)) from MD. 
}
  \label{fig:delta}
\end{figure}

Seeing that the Ubbelohde effect is reproduced in PIMD we now examine the ordering 
behavior at different temperatures.
The proton-ordered AFE phase in squaric acid is characterized by occupation of only one 
site in the H-bonds resulting in long-range order, 
while in the PE phase H/D jumps occur within clusters creating
local defects and disrupting the long-range order. 
A convenient structural order parameter to quantify the ordering is
the $\delta$ parameter, defined as $2\delta=d_{\mathrm{OO}}-2d_{\mathrm{OH}}$ where $d_{\mathrm{OH}}$ 
is the intramolecular O-H (O-D) distance projected onto the OO axis (see Fig.~\ref{fig:vis} bottom 
panel).
Figures~\ref{fig:delta}(a)-(b) reveal characteristics of both AFE and PE ordering in 
probability distributions $P(\delta)$ calculated from PIMD and MD simulations; a unimodal
distribution reflects the absence of jumps while bimodality shows the presence of jumps.
PIMD simulations of H$_2$SQ are ordered at 100~K but disordered at 200~K and above,
while D$_2$SQ is ordered at 100-300~K but disordered at 400-500~K, 
in qualitative agreement with the known difference of 
around 150~K in $T_c$ for the protonated and deuterated crystals. 
On the other hand, 
the MD simulations are ordered at all temperatures, reflecting
the large impact of quantum effects. 
The limited simulation time and system size possible for 
\textit{ab initio} PIMD precludes definitive conclusions on 
AFE or PE ordering in the simulations at a given temperature. However, these two approximations 
can be expected to partially cancel in these rough estimates of $T_c$ since limited simulation time might 
overestimate $T_c$ and insufficiently converged correlations along H-bonding chains should underestimate
$T_c$~\cite{supmat}.
As before, the $P(\delta)$ distributions shown in Figs.~\ref{fig:delta} are calculated along the 
$c$-axis, since 
distributions along the $a$-axis (not shown) are bimodal at all temperatures, even in MD. 

As seen in Figs.~\ref{fig:delta}(a)-(b) H atoms are significantly more likely to populate 
the centric $\delta$=0 position compared to D atoms. Larger probability density at 
$\delta$=0 leads to stronger 
and shorter H-bonds and is the main mechanism behind the Ubbelohde effect, 
as evidenced also by the minimum energy path of proton transfer which involves a shortening of 
$d_{\mathrm{OO}}$ by around 0.15~\AA\ in the $\delta$=0 transition state compared to 
the initial state.
We now examine why there is an enhanced probability at $\delta$=0 in the case of 
H compared to D atoms and in particular consider the role played by tunneling.  

\begin{figure}[h]
\centering
  \includegraphics[width=0.5\textwidth]{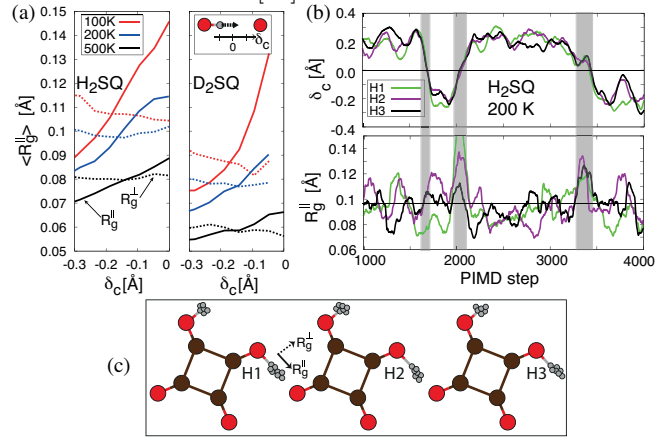}
  \caption{(a) Radius of gyration of H/D ring polymers as function of $\delta_c$ at
    different temperatures for H$_2$SQ (left) and D$_2$SQ (right). Parallel and 
perpendicular components of $R_g$ are shown by solid and dashed lines, respectively. 
(b) Instantaneous H centroid positions (upper) 
and $R_g^{\parallel}$ values (lower) for 3 protons along one H-bonding chain in 
H$_2$SQ at 200~K for a selected portion of the simulation trajectory. Shaded 
areas show collective proton jumps and the horizontal line in the lower panel 
indicates the average $R_g^{\parallel}$. 
(c) Schematic illustration of a collective proton jump and the 
delocalization of H ring-polymers.}
  \label{fig:jumps}
\end{figure}

The role of tunneling is examined by considering the radius of gyration $R_g$ of 
the H and D ring-polymers in our PIMD simulations.
We calculate $R_g$ for a single atom as
$R_g^2=\frac{1}{P}\sum_{i=1}^P (\mathbf{r}_i-\mathbf{r}_c)^2$ 
where the sum runs over the $P$ beads and $\mathbf{r}_c$ is the centroid position 
(center of the ring-polymer). 
$R_g$ thus quantifies the delocalization of a quantum particle, which 
increases during tunneling. 
We find the average values $\left< R_g \right>$ of H atoms to 
decrease from 0.19~\AA\ at 100~K to 0.14~\AA\ at 500~K, and correspondingly 
for D atoms to decrease from 0.15~\AA\ to 0.10~\AA. 
Next we extract the dependence of $R_g$ on $\delta_c$, \textit{i.e.} the $\delta$ parameter for the centroid coordinate, and furthermore decompose into
components parallel ($R_g^{\parallel}$) and perpendicular ($R_g^{\perp}$) 
to the H-bonding vectors (see Fig.~\ref{fig:jumps}(c)).
Figure~\ref{fig:jumps}(a) shows the behavior of  
$R_g^{\parallel}(\delta_c)$ and $R_g^{\perp}(\delta_c)$ for H$_2$SQ 
and D$_2$SQ along the $c$-axis. 
A clear temperature dependence is seen where, in particular, $R_g^{\parallel}$
increases sharply for $\delta_c\rightarrow 0$ at lower temperatures for 
both H$_2$SQ and D$_2$SQ, indicating that tunneling contributes 
significantly when H/D atoms approach the centric 
position of the H-bonds. At higher temperature the increase in $R_g^{\parallel}$ becomes 
smaller but not insignificant, 
indicating that tunneling takes place nearer to the top of the energy barrier due to 
thermal excitations.
It can be seen that the $R_g^{\parallel}$
curve for H$_2$SQ at 100~K extends to $\delta_c$=0  
with a sharp increase in the delocalization even though no collective 
jumps occur at this temperature (see Fig.~\ref{fig:delta}),
showing that protons can tunnel into the central barrier at low temperatures but
quickly return to the original position due to the overall long-range order.
In contrast, $R_g^{\perp}$ is constant or decreases slightly as $\delta_c\rightarrow 0$.
These results suggest that tunneling enhances the probability 
density of H atoms in the centric position and thus contributes directly to the 
Ubbelohde effect.

Tunneling also plays a direct role in collective proton jumps 
as shown in Fig.~\ref{fig:jumps}(b). For each of the 3 protons along 
the $c$-axis we compute the instantaneous values of 
$\delta_c$ and $R_g^{\parallel}$ 
\footnote{No dynamical information can be obtained from PIMD simulations but 
the trajectories sample the quantum mechanical configuration space
and Fig.~\ref{fig:jumps}(b) thus demonstrates real correlations in the system.}.
The upper panel of Fig.~\ref{fig:jumps}(b)
shows several collective proton jumps during a selected portion
of the simulation trajectory for H$_2$SQ at 200~K in one of the sheets and the lower panel 
shows the corresponding instantaneous delocalization.  
In the first shaded collective jump event the instantaneous $R_g^{\parallel}$ of all protons is around the average and the jump thus appears 
to be mostly classical, \textit{i.e.} driven by thermal excitations. In the second event two protons show a large increase in $R_g^{\parallel}$ and 
are clearly tunneling through a part of the energy barrier while the third proton appears to jump by thermal excitation, while in the last event all protons show an increase in delocalization. 
Similar combinations of tunneling and thermal excitations are seen at other temperatures and for D$_2$SQ,
with the probability for non-tunneling thermal jumps increasing at higher temperatures (see 
additional figures in~\cite{supmat}).
To summarize, these results suggest that tunneling of several protons along 
an H-bonding chain not only contributes to the Ubbelohde effect 
but also, in combination with thermal fluctuations, play a role in the 
disordering transition at $T_c$, above which the protons occupy both sites 
in the H-bonds with equal probability.

In conclusion, we find that the combination of dispersion-including (vdW) DFT functionals and
path-integral MD simulations can be used to realistically describe ordered and disordered phases 
of challenging H-bonded ferroelectrics such as squaric acid. A particularly important aspect 
of the simulations is the use of extended supercells that explicitly include correlations between 
H/D atoms along H-bonding chains. 
These correlations help to localize the H/D atoms and without their explicit 
treatment only disordered paraelectric behavior is found.
Our PIMD simulations reveal pronounced nuclear quantum effects where the large 
quantum delocalization of H atoms and tunneling into the central barrier strengthen the H-bonds. 
The same effect is seen in 
the deuterated crystal but with smaller magnitude, leading to an Ubbelohde 
effect with an elongation of oxygen-oxygen distances 
by around 0.02~\AA, in good agreement with experiment.
The Ubbelohde effect in turn leads to 
a large shift in the temperature of the disordering AFE-PE transition upon isotopic 
substitution.
Our simulations also suggest that concerted proton jumps in the 
disordered phase may take place through a combined effect of quantum tunneling of 
several protons combined with thermal fluctuations of nearby non-tunneling protons. 
These insights clarify the role played by tunneling in squaric acid but they 
are likely to extend to other H-bonded ferroelectrics, which 
calls for future simulation studies and experiments. 
The collective jump mechanism assisted by tunneling 
observed here may also contribute to proton transport in, 
\textit{e.g.}, high pressure ice~\cite{schwegler2008melting} 
and along water wires in biological 
environments~\cite{smedarchina2003kinetic} and inside carbon 
nanotubes~\cite{dellago2003proton}.

\subsection*{Acknowledgments}
This research was partially supported by the European Research Council and made 
use of the HECToR supercomputer through membership of the UK's HPC Materials Chemistry Consortium which is funded by the EPSRC (EP/F067496) as well as the 
Abisko supercomputer through the Swedish National Infrastructure for Computing 
(SNIC). 
A.~M. is also supported by the Royal Society through a Royal Society Wolfson Research 
Merit Award, and K.~T.~W. is supported by the Icelandic Research Fund through 
grant no. 120044042.


\begin{thebibliography}{42}%
\makeatletter
\providecommand \@ifxundefined [1]{%
 \@ifx{#1\undefined}
}%
\providecommand \@ifnum [1]{%
 \ifnum #1\expandafter \@firstoftwo
 \else \expandafter \@secondoftwo
 \fi
}%
\providecommand \@ifx [1]{%
 \ifx #1\expandafter \@firstoftwo
 \else \expandafter \@secondoftwo
 \fi
}%
\providecommand \natexlab [1]{#1}%
\providecommand \enquote  [1]{``#1''}%
\providecommand \bibnamefont  [1]{#1}%
\providecommand \bibfnamefont [1]{#1}%
\providecommand \citenamefont [1]{#1}%
\providecommand \href@noop [0]{\@secondoftwo}%
\providecommand \href [0]{\begingroup \@sanitize@url \@href}%
\providecommand \@href[1]{\@@startlink{#1}\@@href}%
\providecommand \@@href[1]{\endgroup#1\@@endlink}%
\providecommand \@sanitize@url [0]{\catcode `\\12\catcode `\$12\catcode
  `\&12\catcode `\#12\catcode `\^12\catcode `\_12\catcode `\%12\relax}%
\providecommand \@@startlink[1]{}%
\providecommand \@@endlink[0]{}%
\providecommand \url  [0]{\begingroup\@sanitize@url \@url }%
\providecommand \@url [1]{\endgroup\@href {#1}{\urlprefix }}%
\providecommand \urlprefix  [0]{URL }%
\providecommand \Eprint [0]{\href }%
\providecommand \doibase [0]{http://dx.doi.org/}%
\providecommand \selectlanguage [0]{\@gobble}%
\providecommand \bibinfo  [0]{\@secondoftwo}%
\providecommand \bibfield  [0]{\@secondoftwo}%
\providecommand \translation [1]{[#1]}%
\providecommand \BibitemOpen [0]{}%
\providecommand \bibitemStop [0]{}%
\providecommand \bibitemNoStop [0]{.\EOS\space}%
\providecommand \EOS [0]{\spacefactor3000\relax}%
\providecommand \BibitemShut  [1]{\csname bibitem#1\endcsname}%
\let\auto@bib@innerbib\@empty
\bibitem [{\citenamefont {Blinc}(2002)}]{blinc2002ferroelectrics}%
  \BibitemOpen
  \bibfield  {author} {\bibinfo {author} {\bibfnamefont {R.}~\bibnamefont
  {Blinc}},\ }\href@noop {} {\bibfield  {journal} {\bibinfo  {journal}
  {Ferroelectrics}\ }\textbf {\bibinfo {volume} {267}},\ \bibinfo {pages} {3}
  (\bibinfo {year} {2002})}\BibitemShut {NoStop}%
\bibitem [{\citenamefont {Horiuchi}\ \emph {et~al.}(2005)\citenamefont
  {Horiuchi}, \citenamefont {Ishii}, \citenamefont {Kumai}, \citenamefont
  {Okimoto}, \citenamefont {Tachibana}, \citenamefont {Nagaosa},\ and\
  \citenamefont {Tokura}}]{horiuchi2005ferro}%
  \BibitemOpen
  \bibfield  {author} {\bibinfo {author} {\bibfnamefont {S.}~\bibnamefont
  {Horiuchi}}, \bibinfo {author} {\bibfnamefont {F.}~\bibnamefont {Ishii}},
  \bibinfo {author} {\bibfnamefont {R.}~\bibnamefont {Kumai}}, \bibinfo
  {author} {\bibfnamefont {Y.}~\bibnamefont {Okimoto}}, \bibinfo {author}
  {\bibfnamefont {H.}~\bibnamefont {Tachibana}}, \bibinfo {author}
  {\bibfnamefont {N.}~\bibnamefont {Nagaosa}}, \ and\ \bibinfo {author}
  {\bibfnamefont {Y.}~\bibnamefont {Tokura}},\ }\href@noop {} {\bibfield
  {journal} {\bibinfo  {journal} {Nature Mater.}\ }\textbf {\bibinfo {volume}
  {4}},\ \bibinfo {pages} {163} (\bibinfo {year} {2005})}\BibitemShut {NoStop}%
\bibitem [{\citenamefont {Horiuchi}\ and\ \citenamefont
  {Tokura}(2008)}]{horiuchi2008organic}%
  \BibitemOpen
  \bibfield  {author} {\bibinfo {author} {\bibfnamefont {S.}~\bibnamefont
  {Horiuchi}}\ and\ \bibinfo {author} {\bibfnamefont {Y.}~\bibnamefont
  {Tokura}},\ }\href@noop {} {\bibfield  {journal} {\bibinfo  {journal} {Nature
  Mater.}\ }\textbf {\bibinfo {volume} {7}},\ \bibinfo {pages} {357} (\bibinfo
  {year} {2008})}\BibitemShut {NoStop}%
\bibitem [{\citenamefont {Horiuchi}\ \emph {et~al.}(2010)\citenamefont
  {Horiuchi}, \citenamefont {Tokunaga}, \citenamefont {Giovannetti},
  \citenamefont {Picozzi}, \citenamefont {Itoh}, \citenamefont {Shimano},
  \citenamefont {Kumai},\ and\ \citenamefont {Tokura}}]{horiuchi2010above}%
  \BibitemOpen
  \bibfield  {author} {\bibinfo {author} {\bibfnamefont {S.}~\bibnamefont
  {Horiuchi}}, \bibinfo {author} {\bibfnamefont {Y.}~\bibnamefont {Tokunaga}},
  \bibinfo {author} {\bibfnamefont {G.}~\bibnamefont {Giovannetti}}, \bibinfo
  {author} {\bibfnamefont {S.}~\bibnamefont {Picozzi}}, \bibinfo {author}
  {\bibfnamefont {H.}~\bibnamefont {Itoh}}, \bibinfo {author} {\bibfnamefont
  {R.}~\bibnamefont {Shimano}}, \bibinfo {author} {\bibfnamefont
  {R.}~\bibnamefont {Kumai}}, \ and\ \bibinfo {author} {\bibfnamefont
  {Y.}~\bibnamefont {Tokura}},\ }\href@noop {} {\bibfield  {journal} {\bibinfo
  {journal} {Nature}\ }\textbf {\bibinfo {volume} {463}},\ \bibinfo {pages}
  {789} (\bibinfo {year} {2010})}\BibitemShut {NoStop}%
\bibitem [{\citenamefont {Blinc}(1960)}]{blinc1960isotopic}%
  \BibitemOpen
  \bibfield  {author} {\bibinfo {author} {\bibfnamefont {R.}~\bibnamefont
  {Blinc}},\ }\href@noop {} {\bibfield  {journal} {\bibinfo  {journal} {J.
  Phys. Chem. Solids}\ }\textbf {\bibinfo {volume} {13}},\ \bibinfo {pages}
  {204} (\bibinfo {year} {1960})}\BibitemShut {NoStop}%
\bibitem [{\citenamefont {Blinc}\ and\ \citenamefont
  {Svetina}(1966)}]{blinc1966cluster1}%
  \BibitemOpen
  \bibfield  {author} {\bibinfo {author} {\bibfnamefont {R.}~\bibnamefont
  {Blinc}}\ and\ \bibinfo {author} {\bibfnamefont {S.}~\bibnamefont
  {Svetina}},\ }\href@noop {} {\bibfield  {journal} {\bibinfo  {journal} {Phys.
  Rev.}\ }\textbf {\bibinfo {volume} {147}},\ \bibinfo {pages} {423} (\bibinfo
  {year} {1966})}\BibitemShut {NoStop}%
\bibitem [{\citenamefont {McMahon}\ \emph {et~al.}(1990)\citenamefont
  {McMahon}, \citenamefont {Nelmes}, \citenamefont {Kuhst}, \citenamefont
  {Dorwarth}, \citenamefont {Piltz},\ and\ \citenamefont
  {Tun}}]{mcmahon1990geometric}%
  \BibitemOpen
  \bibfield  {author} {\bibinfo {author} {\bibfnamefont {M.~I.}\ \bibnamefont
  {McMahon}}, \bibinfo {author} {\bibfnamefont {R.~J.}\ \bibnamefont {Nelmes}},
  \bibinfo {author} {\bibfnamefont {W.~F.}\ \bibnamefont {Kuhst}}, \bibinfo
  {author} {\bibfnamefont {R.}~\bibnamefont {Dorwarth}}, \bibinfo {author}
  {\bibfnamefont {R.~O.}\ \bibnamefont {Piltz}}, \ and\ \bibinfo {author}
  {\bibfnamefont {Z.}~\bibnamefont {Tun}},\ }\href@noop {} {\bibfield
  {journal} {\bibinfo  {journal} {Nature}\ }\textbf {\bibinfo {volume} {348}},\
  \bibinfo {pages} {317} (\bibinfo {year} {1990})}\BibitemShut {NoStop}%
\bibitem [{\citenamefont {McMahon}\ \emph {et~al.}(1991)\citenamefont
  {McMahon}, \citenamefont {Nelmes}, \citenamefont {Kuhs},\ and\ \citenamefont
  {Semmingsen}}]{mcmahon1991effect}%
  \BibitemOpen
  \bibfield  {author} {\bibinfo {author} {\bibfnamefont {M.~I.}\ \bibnamefont
  {McMahon}}, \bibinfo {author} {\bibfnamefont {R.~J.}\ \bibnamefont {Nelmes}},
  \bibinfo {author} {\bibfnamefont {W.~F.}\ \bibnamefont {Kuhs}}, \ and\
  \bibinfo {author} {\bibfnamefont {D.}~\bibnamefont {Semmingsen}},\
  }\href@noop {} {\bibfield  {journal} {\bibinfo  {journal} {Z. f.
  Kristallogr.}\ }\textbf {\bibinfo {volume} {195}},\ \bibinfo {pages} {231}
  (\bibinfo {year} {1991})}\BibitemShut {NoStop}%
\bibitem [{\citenamefont {Ubbelohde}\ and\ \citenamefont
  {Gallagher}(1955)}]{ubbelohde1955acid}%
  \BibitemOpen
  \bibfield  {author} {\bibinfo {author} {\bibfnamefont {A.~R.}\ \bibnamefont
  {Ubbelohde}}\ and\ \bibinfo {author} {\bibfnamefont {K.~J.}\ \bibnamefont
  {Gallagher}},\ }\href@noop {} {\bibfield  {journal} {\bibinfo  {journal}
  {Acta Crystallogr.}\ }\textbf {\bibinfo {volume} {8}},\ \bibinfo {pages} {71}
  (\bibinfo {year} {1955})}\BibitemShut {NoStop}%
\bibitem [{\citenamefont {Li}\ \emph {et~al.}(2011)\citenamefont {Li},
  \citenamefont {Walker},\ and\ \citenamefont {Michaelides}}]{li2011quantum}%
  \BibitemOpen
  \bibfield  {author} {\bibinfo {author} {\bibfnamefont {X.~Z.}\ \bibnamefont
  {Li}}, \bibinfo {author} {\bibfnamefont {B.}~\bibnamefont {Walker}}, \ and\
  \bibinfo {author} {\bibfnamefont {A.}~\bibnamefont {Michaelides}},\
  }\href@noop {} {\bibfield  {journal} {\bibinfo  {journal} {Proc. Natl. Acad.
  Sci. (USA)}\ }\textbf {\bibinfo {volume} {108}},\ \bibinfo {pages} {6369}
  (\bibinfo {year} {2011})}\BibitemShut {NoStop}%
\bibitem [{\citenamefont {Bussmann-Holder}\ and\ \citenamefont
  {Michel}(1998)}]{bussmann1998bond}%
  \BibitemOpen
  \bibfield  {author} {\bibinfo {author} {\bibfnamefont {A.}~\bibnamefont
  {Bussmann-Holder}}\ and\ \bibinfo {author} {\bibfnamefont {K.~H.}\
  \bibnamefont {Michel}},\ }\href@noop {} {\bibfield  {journal} {\bibinfo
  {journal} {Phys. Rev. Lett.}\ }\textbf {\bibinfo {volume} {80}},\ \bibinfo
  {pages} {2173} (\bibinfo {year} {1998})}\BibitemShut {NoStop}%
\bibitem [{\citenamefont {Dalal}\ \emph {et~al.}(1998)\citenamefont {Dalal},
  \citenamefont {Klymachyov},\ and\ \citenamefont
  {Bussmann-Holder}}]{dalal1998coexistence}%
  \BibitemOpen
  \bibfield  {author} {\bibinfo {author} {\bibfnamefont {N.}~\bibnamefont
  {Dalal}}, \bibinfo {author} {\bibfnamefont {A.}~\bibnamefont {Klymachyov}}, \
  and\ \bibinfo {author} {\bibfnamefont {A.}~\bibnamefont {Bussmann-Holder}},\
  }\href@noop {} {\bibfield  {journal} {\bibinfo  {journal} {Phys. Rev. Lett.}\
  }\textbf {\bibinfo {volume} {81}},\ \bibinfo {pages} {5924} (\bibinfo {year}
  {1998})}\BibitemShut {NoStop}%
\bibitem [{\citenamefont {Reiter}\ \emph {et~al.}(2002)\citenamefont {Reiter},
  \citenamefont {Mayers},\ and\ \citenamefont {Platzman}}]{reiter2002direct}%
  \BibitemOpen
  \bibfield  {author} {\bibinfo {author} {\bibfnamefont {G.~F.}\ \bibnamefont
  {Reiter}}, \bibinfo {author} {\bibfnamefont {J.}~\bibnamefont {Mayers}}, \
  and\ \bibinfo {author} {\bibfnamefont {P.}~\bibnamefont {Platzman}},\
  }\href@noop {} {\bibfield  {journal} {\bibinfo  {journal} {Phys. Rev. Lett.}\
  }\textbf {\bibinfo {volume} {89}},\ \bibinfo {pages} {135505} (\bibinfo
  {year} {2002})}\BibitemShut {NoStop}%
\bibitem [{\citenamefont {Reiter}\ \emph {et~al.}(2008)\citenamefont {Reiter},
  \citenamefont {Shukla}, \citenamefont {Platzman},\ and\ \citenamefont
  {Mayers}}]{reiter2008deuteron}%
  \BibitemOpen
  \bibfield  {author} {\bibinfo {author} {\bibfnamefont {G.}~\bibnamefont
  {Reiter}}, \bibinfo {author} {\bibfnamefont {A.}~\bibnamefont {Shukla}},
  \bibinfo {author} {\bibfnamefont {P.}~\bibnamefont {Platzman}}, \ and\
  \bibinfo {author} {\bibfnamefont {J.}~\bibnamefont {Mayers}},\ }\href@noop {}
  {\bibfield  {journal} {\bibinfo  {journal} {New J. Phys.}\ }\textbf {\bibinfo
  {volume} {10}},\ \bibinfo {pages} {013016} (\bibinfo {year}
  {2008})}\BibitemShut {NoStop}%
\bibitem [{\citenamefont {Koval}\ \emph {et~al.}(2002)\citenamefont {Koval},
  \citenamefont {Kohanoff}, \citenamefont {Migoni},\ and\ \citenamefont
  {Tosatti}}]{koval2002ferroelectricity}%
  \BibitemOpen
  \bibfield  {author} {\bibinfo {author} {\bibfnamefont {S.}~\bibnamefont
  {Koval}}, \bibinfo {author} {\bibfnamefont {J.}~\bibnamefont {Kohanoff}},
  \bibinfo {author} {\bibfnamefont {R.~L.}\ \bibnamefont {Migoni}}, \ and\
  \bibinfo {author} {\bibfnamefont {E.}~\bibnamefont {Tosatti}},\ }\href@noop
  {} {\bibfield  {journal} {\bibinfo  {journal} {Phys. Rev. Lett.}\ }\textbf
  {\bibinfo {volume} {89}},\ \bibinfo {pages} {187602} (\bibinfo {year}
  {2002})}\BibitemShut {NoStop}%
\bibitem [{\citenamefont {Koval}\ \emph {et~al.}(2005)\citenamefont {Koval},
  \citenamefont {Kohanoff}, \citenamefont {Lasave}, \citenamefont {Colizzi},\
  and\ \citenamefont {Migoni}}]{koval2005first}%
  \BibitemOpen
  \bibfield  {author} {\bibinfo {author} {\bibfnamefont {S.}~\bibnamefont
  {Koval}}, \bibinfo {author} {\bibfnamefont {J.}~\bibnamefont {Kohanoff}},
  \bibinfo {author} {\bibfnamefont {J.}~\bibnamefont {Lasave}}, \bibinfo
  {author} {\bibfnamefont {G.}~\bibnamefont {Colizzi}}, \ and\ \bibinfo
  {author} {\bibfnamefont {R.~L.}\ \bibnamefont {Migoni}},\ }\href@noop {}
  {\bibfield  {journal} {\bibinfo  {journal} {Phys. Rev. B}\ }\textbf {\bibinfo
  {volume} {71}},\ \bibinfo {pages} {184102} (\bibinfo {year}
  {2005})}\BibitemShut {NoStop}%
\bibitem [{\citenamefont {Petersson}(1981)}]{petersson1981phase}%
  \BibitemOpen
  \bibfield  {author} {\bibinfo {author} {\bibfnamefont {J.}~\bibnamefont
  {Petersson}},\ }\href@noop {} {\bibfield  {journal} {\bibinfo  {journal}
  {Ferroelectrics}\ }\textbf {\bibinfo {volume} {35}},\ \bibinfo {pages} {57}
  (\bibinfo {year} {1981})}\BibitemShut {NoStop}%
\bibitem [{\citenamefont {Katrusiak}\ and\ \citenamefont
  {Nelmes}(1986)}]{katrusiak1986pressure}%
  \BibitemOpen
  \bibfield  {author} {\bibinfo {author} {\bibfnamefont {A.}~\bibnamefont
  {Katrusiak}}\ and\ \bibinfo {author} {\bibfnamefont {R.}~\bibnamefont
  {Nelmes}},\ }\href@noop {} {\bibfield  {journal} {\bibinfo  {journal} {J.
  Phys. C: Solid State Phys.}\ }\textbf {\bibinfo {volume} {19}},\ \bibinfo
  {pages} {L765} (\bibinfo {year} {1986})}\BibitemShut {NoStop}%
\bibitem [{\citenamefont {Semmingsen}\ \emph {et~al.}(1995)\citenamefont
  {Semmingsen}, \citenamefont {Tun}, \citenamefont {Nelmes}, \citenamefont
  {McMullan},\ and\ \citenamefont {Koetzle}}]{semmingsen1995ZfK}%
  \BibitemOpen
  \bibfield  {author} {\bibinfo {author} {\bibfnamefont {D.}~\bibnamefont
  {Semmingsen}}, \bibinfo {author} {\bibfnamefont {Z.}~\bibnamefont {Tun}},
  \bibinfo {author} {\bibfnamefont {R.~J.}\ \bibnamefont {Nelmes}}, \bibinfo
  {author} {\bibfnamefont {R.~K.}\ \bibnamefont {McMullan}}, \ and\ \bibinfo
  {author} {\bibfnamefont {T.~F.}\ \bibnamefont {Koetzle}},\ }\href@noop {}
  {\bibfield  {journal} {\bibinfo  {journal} {Z. f. Kristallogr.}\ }\textbf
  {\bibinfo {volume} {210}},\ \bibinfo {pages} {934} (\bibinfo {year}
  {1995})}\BibitemShut {NoStop}%
\bibitem [{\citenamefont {Dolin}\ \emph {et~al.}(2011)\citenamefont {Dolin},
  \citenamefont {Levin}, \citenamefont {Mikhailova}, \citenamefont {Solin},\
  and\ \citenamefont {Zinova}}]{dolin2011quantum}%
  \BibitemOpen
  \bibfield  {author} {\bibinfo {author} {\bibfnamefont {S.}~\bibnamefont
  {Dolin}}, \bibinfo {author} {\bibfnamefont {A.}~\bibnamefont {Levin}},
  \bibinfo {author} {\bibfnamefont {T.~Y.}\ \bibnamefont {Mikhailova}},
  \bibinfo {author} {\bibfnamefont {M.}~\bibnamefont {Solin}}, \ and\ \bibinfo
  {author} {\bibfnamefont {N.}~\bibnamefont {Zinova}},\ }\href@noop {}
  {\bibfield  {journal} {\bibinfo  {journal} {Int. J. Quant. Chem.}\ }\textbf
  {\bibinfo {volume} {111}},\ \bibinfo {pages} {2671} (\bibinfo {year}
  {2011})}\BibitemShut {NoStop}%
\bibitem [{\citenamefont {Rovira}\ \emph {et~al.}(2001)\citenamefont {Rovira},
  \citenamefont {Novoa},\ and\ \citenamefont {Ballone}}]{rovira2001hydrogen}%
  \BibitemOpen
  \bibfield  {author} {\bibinfo {author} {\bibfnamefont {C.}~\bibnamefont
  {Rovira}}, \bibinfo {author} {\bibfnamefont {J.~J.}\ \bibnamefont {Novoa}}, \
  and\ \bibinfo {author} {\bibfnamefont {P.}~\bibnamefont {Ballone}},\
  }\href@noop {} {\bibfield  {journal} {\bibinfo  {journal} {J. Chem. Phys.}\
  }\textbf {\bibinfo {volume} {115}},\ \bibinfo {pages} {6406} (\bibinfo {year}
  {2001})}\BibitemShut {NoStop}%
\bibitem [{\citenamefont {Palomar}\ and\ \citenamefont
  {Dalal}(2002{\natexlab{a}})}]{palomar2002GIAO}%
  \BibitemOpen
  \bibfield  {author} {\bibinfo {author} {\bibfnamefont {J.}~\bibnamefont
  {Palomar}}\ and\ \bibinfo {author} {\bibfnamefont {N.~S.}\ \bibnamefont
  {Dalal}},\ }\href@noop {} {\bibfield  {journal} {\bibinfo  {journal}
  {Ferroelectrics}\ }\textbf {\bibinfo {volume} {272}},\ \bibinfo {pages} {173}
  (\bibinfo {year} {2002}{\natexlab{a}})}\BibitemShut {NoStop}%
\bibitem [{\citenamefont {Bussmann-Holder}\ and\ \citenamefont
  {Dalal}(2007)}]{bussmann2007order}%
  \BibitemOpen
  \bibfield  {author} {\bibinfo {author} {\bibfnamefont {A.}~\bibnamefont
  {Bussmann-Holder}}\ and\ \bibinfo {author} {\bibfnamefont {N.}~\bibnamefont
  {Dalal}},\ }in\ \href@noop {} {\emph {\bibinfo {booktitle} {Struct. Bond.}}}\
  (\bibinfo  {publisher} {Springer},\ \bibinfo {year} {2007})\ pp.\ \bibinfo
  {pages} {1--21}\BibitemShut {NoStop}%
\bibitem [{\citenamefont {Ishizuka}\ \emph {et~al.}(2011)\citenamefont
  {Ishizuka}, \citenamefont {Motome}, \citenamefont {Furukawa},\ and\
  \citenamefont {Suzuki}}]{ishizuka2011quantum}%
  \BibitemOpen
  \bibfield  {author} {\bibinfo {author} {\bibfnamefont {H.}~\bibnamefont
  {Ishizuka}}, \bibinfo {author} {\bibfnamefont {Y.}~\bibnamefont {Motome}},
  \bibinfo {author} {\bibfnamefont {N.}~\bibnamefont {Furukawa}}, \ and\
  \bibinfo {author} {\bibfnamefont {S.}~\bibnamefont {Suzuki}},\ }\href@noop {}
  {\bibfield  {journal} {\bibinfo  {journal} {Phys. Rev. B}\ }\textbf {\bibinfo
  {volume} {84}},\ \bibinfo {pages} {064120} (\bibinfo {year}
  {2011})}\BibitemShut {NoStop}%
\bibitem [{\citenamefont {Srinivasan}\ and\ \citenamefont
  {Sebastiani}(2011)}]{srinivasan2011isotope}%
  \BibitemOpen
  \bibfield  {author} {\bibinfo {author} {\bibfnamefont {V.}~\bibnamefont
  {Srinivasan}}\ and\ \bibinfo {author} {\bibfnamefont {D.}~\bibnamefont
  {Sebastiani}},\ }\href@noop {} {\bibfield  {journal} {\bibinfo  {journal} {J.
  Phys. Chem. C}\ }\textbf {\bibinfo {volume} {115}},\ \bibinfo {pages} {12631}
  (\bibinfo {year} {2011})}\BibitemShut {NoStop}%
\bibitem [{\citenamefont {{S}ee {S}upplemental~{M}aterial at:}()}]{supmat}%
  \BibitemOpen
  \bibfield  {author} {\bibinfo {author} {\bibnamefont {{S}ee
  {S}upplemental~{M}aterial at:}},\ }\href {http://www.inserturlhere.com}
  {}\BibitemShut {NoStop}%
\bibitem [{\citenamefont {Kresse}\ and\ \citenamefont
  {Joubert}(1999)}]{kresse1999ultrasoft}%
  \BibitemOpen
  \bibfield  {author} {\bibinfo {author} {\bibfnamefont {G.}~\bibnamefont
  {Kresse}}\ and\ \bibinfo {author} {\bibfnamefont {D.}~\bibnamefont
  {Joubert}},\ }\href@noop {} {\bibfield  {journal} {\bibinfo  {journal} {Phys.
  Rev. B}\ }\textbf {\bibinfo {volume} {59}},\ \bibinfo {pages} {1758}
  (\bibinfo {year} {1999})}\BibitemShut {NoStop}%
\bibitem [{\citenamefont {Kresse}\ and\ \citenamefont
  {Furthm\"uller}(1996{\natexlab{a}})}]{kresse1996vasp1}%
  \BibitemOpen
  \bibfield  {author} {\bibinfo {author} {\bibfnamefont {G.}~\bibnamefont
  {Kresse}}\ and\ \bibinfo {author} {\bibfnamefont {J.}~\bibnamefont
  {Furthm\"uller}},\ }\href@noop {} {\bibfield  {journal} {\bibinfo  {journal}
  {J. Comp. Mater. Sci.}\ }\textbf {\bibinfo {volume} {6}},\ \bibinfo {pages}
  {15} (\bibinfo {year} {1996}{\natexlab{a}})}\BibitemShut {NoStop}%
\bibitem [{\citenamefont {Kresse}\ and\ \citenamefont
  {Furthm\"uller}(1996{\natexlab{b}})}]{kresse1996vasp2}%
  \BibitemOpen
  \bibfield  {author} {\bibinfo {author} {\bibfnamefont {G.}~\bibnamefont
  {Kresse}}\ and\ \bibinfo {author} {\bibfnamefont {J.}~\bibnamefont
  {Furthm\"uller}},\ }\href@noop {} {\bibfield  {journal} {\bibinfo  {journal}
  {Phys. Rev. B}\ }\textbf {\bibinfo {volume} {54}},\ \bibinfo {pages} {11169}
  (\bibinfo {year} {1996}{\natexlab{b}})}\BibitemShut {NoStop}%
\bibitem [{\citenamefont {Marx}\ and\ \citenamefont
  {Parrinello}(1994)}]{marx1994ab}%
  \BibitemOpen
  \bibfield  {author} {\bibinfo {author} {\bibfnamefont {D.}~\bibnamefont
  {Marx}}\ and\ \bibinfo {author} {\bibfnamefont {M.}~\bibnamefont
  {Parrinello}},\ }\href@noop {} {\bibfield  {journal} {\bibinfo  {journal} {Z.
  f. Phys. B: Cond. Matt.}\ }\textbf {\bibinfo {volume} {95}},\ \bibinfo
  {pages} {143} (\bibinfo {year} {1994})}\BibitemShut {NoStop}%
\bibitem [{\citenamefont {Tuckerman}\ \emph {et~al.}(1997)\citenamefont
  {Tuckerman}, \citenamefont {Marx}, \citenamefont {Klein},\ and\ \citenamefont
  {Parrinello}}]{tuckerman1997quantum}%
  \BibitemOpen
  \bibfield  {author} {\bibinfo {author} {\bibfnamefont {M.~E.}\ \bibnamefont
  {Tuckerman}}, \bibinfo {author} {\bibfnamefont {D.}~\bibnamefont {Marx}},
  \bibinfo {author} {\bibfnamefont {M.~L.}\ \bibnamefont {Klein}}, \ and\
  \bibinfo {author} {\bibfnamefont {M.}~\bibnamefont {Parrinello}},\
  }\href@noop {} {\bibfield  {journal} {\bibinfo  {journal} {Science}\ }\textbf
  {\bibinfo {volume} {275}},\ \bibinfo {pages} {817} (\bibinfo {year}
  {1997})}\BibitemShut {NoStop}%
\bibitem [{\citenamefont {Alf{\`e}}\ and\ \citenamefont
  {Gillan}(2010)}]{alfe2010ab}%
  \BibitemOpen
  \bibfield  {author} {\bibinfo {author} {\bibfnamefont {D.}~\bibnamefont
  {Alf{\`e}}}\ and\ \bibinfo {author} {\bibfnamefont {M.~J.}\ \bibnamefont
  {Gillan}},\ }\href@noop {} {\bibfield  {journal} {\bibinfo  {journal} {J.
  Chem. Phys.}\ }\textbf {\bibinfo {volume} {133}},\ \bibinfo {pages} {044103}
  (\bibinfo {year} {2010})}\BibitemShut {NoStop}%
\bibitem [{\citenamefont {Ehrhardt}\ \emph {et~al.}(1984)\citenamefont
  {Ehrhardt}, \citenamefont {Buchenau}, \citenamefont {Samuelsen},\ and\
  \citenamefont {Maier}}]{ehrhardt1984one}%
  \BibitemOpen
  \bibfield  {author} {\bibinfo {author} {\bibfnamefont {K.-D.}\ \bibnamefont
  {Ehrhardt}}, \bibinfo {author} {\bibfnamefont {U.}~\bibnamefont {Buchenau}},
  \bibinfo {author} {\bibfnamefont {E.}~\bibnamefont {Samuelsen}}, \ and\
  \bibinfo {author} {\bibfnamefont {H.}~\bibnamefont {Maier}},\ }\href@noop {}
  {\bibfield  {journal} {\bibinfo  {journal} {Phys. Rev. B}\ }\textbf {\bibinfo
  {volume} {29}},\ \bibinfo {pages} {996} (\bibinfo {year} {1984})}\BibitemShut
  {NoStop}%
\bibitem [{\citenamefont {Palomar}\ and\ \citenamefont
  {Dalal}(2002{\natexlab{b}})}]{palomar2002quantum}%
  \BibitemOpen
  \bibfield  {author} {\bibinfo {author} {\bibfnamefont {J.}~\bibnamefont
  {Palomar}}\ and\ \bibinfo {author} {\bibfnamefont {N.}~\bibnamefont
  {Dalal}},\ }\href@noop {} {\bibfield  {journal} {\bibinfo  {journal} {J.
  Phys. Chem. B}\ }\textbf {\bibinfo {volume} {106}},\ \bibinfo {pages} {4799}
  (\bibinfo {year} {2002}{\natexlab{b}})}\BibitemShut {NoStop}%
\bibitem [{\citenamefont {Lee}\ \emph {et~al.}(2010)\citenamefont {Lee},
  \citenamefont {Murray}, \citenamefont {Kong}, \citenamefont {Lundqvist},\
  and\ \citenamefont {Langreth}}]{lee2010vdwdf2}%
  \BibitemOpen
  \bibfield  {author} {\bibinfo {author} {\bibfnamefont {K.}~\bibnamefont
  {Lee}}, \bibinfo {author} {\bibfnamefont {{\'E}.~D.}\ \bibnamefont {Murray}},
  \bibinfo {author} {\bibfnamefont {L.}~\bibnamefont {Kong}}, \bibinfo {author}
  {\bibfnamefont {B.~I.}\ \bibnamefont {Lundqvist}}, \ and\ \bibinfo {author}
  {\bibfnamefont {D.~C.}\ \bibnamefont {Langreth}},\ }\href {\doibase
  10.1103/PhysRevB.82.081101} {\bibfield  {journal} {\bibinfo  {journal} {Phys.
  Rev. B}\ }\textbf {\bibinfo {volume} {82}},\ \bibinfo {pages} {081101}
  (\bibinfo {year} {2010})}\BibitemShut {NoStop}%
\bibitem [{\citenamefont {Klime{\v{s}}}\ \emph {et~al.}(2011)\citenamefont
  {Klime{\v{s}}}, \citenamefont {Bowler},\ and\ \citenamefont
  {Michaelides}}]{klimes2011van}%
  \BibitemOpen
  \bibfield  {author} {\bibinfo {author} {\bibfnamefont {J.}~\bibnamefont
  {Klime{\v{s}}}}, \bibinfo {author} {\bibfnamefont {D.~R.}\ \bibnamefont
  {Bowler}}, \ and\ \bibinfo {author} {\bibfnamefont {A.}~\bibnamefont
  {Michaelides}},\ }\href@noop {} {\bibfield  {journal} {\bibinfo  {journal}
  {Phys. Rev. B}\ }\textbf {\bibinfo {volume} {83}},\ \bibinfo {pages} {195131}
  (\bibinfo {year} {2011})}\BibitemShut {NoStop}%
\bibitem [{\citenamefont {Semmingsen}\ \emph {et~al.}(1977)\citenamefont
  {Semmingsen}, \citenamefont {Hollander},\ and\ \citenamefont
  {Koetzle}}]{semmingsen1977jcp}%
  \BibitemOpen
  \bibfield  {author} {\bibinfo {author} {\bibfnamefont {D.}~\bibnamefont
  {Semmingsen}}, \bibinfo {author} {\bibfnamefont {F.~J.}\ \bibnamefont
  {Hollander}}, \ and\ \bibinfo {author} {\bibfnamefont {T.~F.}\ \bibnamefont
  {Koetzle}},\ }\href@noop {} {\bibfield  {journal} {\bibinfo  {journal} {J.
  Chem. Phys.}\ }\textbf {\bibinfo {volume} {66}},\ \bibinfo {pages} {4405}
  (\bibinfo {year} {1977})}\BibitemShut {NoStop}%
\bibitem [{\citenamefont {Lee}\ \emph {et~al.}(2012)\citenamefont {Lee},
  \citenamefont {Jung~Kweon}, \citenamefont {Oh},\ and\ \citenamefont
  {Eui~Lee}}]{lee2012polymorphic}%
  \BibitemOpen
  \bibfield  {author} {\bibinfo {author} {\bibfnamefont {K.-S.}\ \bibnamefont
  {Lee}}, \bibinfo {author} {\bibfnamefont {J.}~\bibnamefont {Jung~Kweon}},
  \bibinfo {author} {\bibfnamefont {I.-H.}\ \bibnamefont {Oh}}, \ and\ \bibinfo
  {author} {\bibfnamefont {C.}~\bibnamefont {Eui~Lee}},\ }\href@noop {}
  {\bibfield  {journal} {\bibinfo  {journal} {J. Phys. Chem. Solids}\ }\textbf
  {\bibinfo {volume} {73}},\ \bibinfo {pages} {890} (\bibinfo {year}
  {2012})}\BibitemShut {NoStop}%
\bibitem [{Note1()}]{Note1}%
  \BibitemOpen
  \bibinfo {note} {No dynamical information can be obtained from PIMD
  simulations but the trajectories sample the quantum mechanical configuration
  space and Fig.~\ref {fig:jumps}(b) thus demonstrates real correlations in the
  system.}\BibitemShut {Stop}%
\bibitem [{\citenamefont {Schwegler}\ \emph {et~al.}(2008)\citenamefont
  {Schwegler}, \citenamefont {Sharma}, \citenamefont {Gygi},\ and\
  \citenamefont {Galli}}]{schwegler2008melting}%
  \BibitemOpen
  \bibfield  {author} {\bibinfo {author} {\bibfnamefont {E.}~\bibnamefont
  {Schwegler}}, \bibinfo {author} {\bibfnamefont {M.}~\bibnamefont {Sharma}},
  \bibinfo {author} {\bibfnamefont {F.}~\bibnamefont {Gygi}}, \ and\ \bibinfo
  {author} {\bibfnamefont {G.}~\bibnamefont {Galli}},\ }\href@noop {}
  {\bibfield  {journal} {\bibinfo  {journal} {Proc. Natl. Acad. Sci. (USA)}\
  }\textbf {\bibinfo {volume} {105}},\ \bibinfo {pages} {14779} (\bibinfo
  {year} {2008})}\BibitemShut {NoStop}%
\bibitem [{\citenamefont {Smedarchina}\ \emph {et~al.}(2003)\citenamefont
  {Smedarchina}, \citenamefont {Siebrand}, \citenamefont
  {Fern{\'a}ndez-Ramos},\ and\ \citenamefont {Cui}}]{smedarchina2003kinetic}%
  \BibitemOpen
  \bibfield  {author} {\bibinfo {author} {\bibfnamefont {Z.}~\bibnamefont
  {Smedarchina}}, \bibinfo {author} {\bibfnamefont {W.}~\bibnamefont
  {Siebrand}}, \bibinfo {author} {\bibfnamefont {A.}~\bibnamefont
  {Fern{\'a}ndez-Ramos}}, \ and\ \bibinfo {author} {\bibfnamefont
  {Q.}~\bibnamefont {Cui}},\ }\href@noop {} {\bibfield  {journal} {\bibinfo
  {journal} {J. Am. Chem. Soc.}\ }\textbf {\bibinfo {volume} {125}},\ \bibinfo
  {pages} {243} (\bibinfo {year} {2003})}\BibitemShut {NoStop}%
\bibitem [{\citenamefont {Dellago}\ \emph {et~al.}(2003)\citenamefont
  {Dellago}, \citenamefont {Naor},\ and\ \citenamefont
  {Hummer}}]{dellago2003proton}%
  \BibitemOpen
  \bibfield  {author} {\bibinfo {author} {\bibfnamefont {C.}~\bibnamefont
  {Dellago}}, \bibinfo {author} {\bibfnamefont {M.~M.}\ \bibnamefont {Naor}}, \
  and\ \bibinfo {author} {\bibfnamefont {G.}~\bibnamefont {Hummer}},\
  }\href@noop {} {\bibfield  {journal} {\bibinfo  {journal} {Phys. Rev. Lett.}\
  }\textbf {\bibinfo {volume} {90}},\ \bibinfo {pages} {105902} (\bibinfo
  {year} {2003})}\BibitemShut {NoStop}%
\end{thebibliography}

%

\end{document}